\documentclass[12pt]{article}
\usepackage{epsfig}
\usepackage{amsfonts}
\usepackage{latexsym}
\usepackage{amsmath}
\usepackage{amssymb}
\usepackage{mathrsfs}
\usepackage{hyperref}
\usepackage{cite}
\numberwithin{equation}{section}
			    
\DeclareFontFamily{OT1}{rsfs}{}
\DeclareFontShape{OT1}{rsfs}{m}{n}{
<-7> rsfs5 <7-10> rsfs7 <10-> rsfs10}{}
\DeclareMathAlphabet{\mycal}{OT1}{rsfs}{m}{n}

\def\half{{1\over 2}}
\newcommand{\p}{\partial}

\newcommand{\bea}{\begin{eqnarray}}
\newcommand{\eea}{\end{eqnarray}}
\newcommand{\beas}{\begin{eqnarray*}}
\newcommand{\eeas}{\end{eqnarray*}}
\newcommand{\be}{\begin{equation}}
\newcommand{\ee}{\end{equation}}

  \makeatletter
  \let\over=\@@over \let\overwithdelims=\@@overwithdelims
  \let\atop=\@@atop \let\atopwithdelims=\@@atopwithdelims
  \let\above=\@@above \let\abovewithdelims=\@@abovewithdelims
  \makeatother

\begin{document}

\begin{titlepage}
\unitlength = 1mm
\begin{center}
\vskip1cm
{ \LARGE {\textsc{Photon Emission Near Extreme }}}
\vskip.3cm
{ \LARGE {\textsc{Kerr Black Holes}}}

\vspace{0.8cm}
Achilleas P. Porfyriadis, Yichen Shi and Andrew Strominger

\vspace{.5cm}

{\it  Center for the Fundamental Laws of Nature, Harvard University,\\
Cambridge, MA 02138, USA}
\vskip2cm
\begin{abstract}
Ongoing astronomical efforts extract physical properties of black holes from electromagnetic emissions in their near-vicinity. This requires finding the null geodesics which extend from the near-horizon region out to a distant observatory. In general these can only be found numerically.  In this paper, for the interesting special case of extremally spinning Kerr black holes, we use an emergent near-horizon conformal symmetry to find near-superradiant geodesics analytically.

\end{abstract}

\vspace{1.0cm}

\end{center}
\end{titlepage}

\pagestyle{plain}
\setcounter{page}{1}
\newcounter{bean}
\baselineskip18pt


\setcounter{tocdepth}{2}
\section{Introduction}
In the coming decades,  LIGO \cite{ligo}, the Event Horizon Telescope \cite{eht} and a number of other experiments \cite{athena,ska,elisa} are expected to advance observational black hole 
astrophysics to a qualitatively new level of precision. The phenomenology of black holes is based on the Kerr solution which describes neutral rotating black holes in vacuum general relativity. A large number of observed black holes, especially supermassive ones in active galactic nuclei, are thought to be rotating very fast \cite{Reynolds:2013qqa,Brenneman:2013oba}. Therefore extreme Kerr, which is rotating at the maximally allowed theoretical limit, is of particular observational relevance.

From a purely theoretical perspective extreme Kerr is also of particular interest  for several reasons. An enhanced conformal symmetry group of the geometry emerges as the horizon is approached \cite{Bardeen:1999px, Guica:2008mu} which greatly constrains and simplifies the gravitational dynamics in the near-horizon region. This is known as the Kerr/CFT (Conformal Field Theory) correspondence, because the same symmetry group appears in two-dimensional CFTs. The conformal symmetries imply a variety of qualitatively new and potentially observable phenomena \cite{Porfyriadis:2014fja,Hadar:2014dpa,Lupsasca:2014pfa, diracmedaltalk,Lupsasca:2014hua,Hadar:2015xpa,Gralla:2015rpa,Compere:2015pja,Gralla:2016jfc,Gralla:2016qfw,Casals:2016mel}. At the quantum level, they have conjectural implications for black hole information \cite{Guica:2008mu,Bredberg:2011hp,Compere:2012jk}. More generally, the extreme Kerr black hole is a critical point in the Kerr family of astrophysical black holes and hence worthy of special attention. 

Recently the Kerr/CFT correspondence has been used to compute explicit analytical gravitational waveforms for extreme-mass-ratio-inspirals in the near-horizon region of near-extreme Kerr black holes \cite{Porfyriadis:2014fja,Hadar:2014dpa,Hadar:2015xpa,Gralla:2015rpa}. It has also been used to analytically solve the equations of force-free electrodynamics which are expected to govern the near-horizon magnetosphere \cite{Lupsasca:2014pfa,Lupsasca:2014hua,Compere:2015pja,Gralla:2016jfc}. 

In this paper we lay the groundwork for an exploration of the consequences of Kerr/CFT on electromagnetic emissions from the near-horizon region of an extreme Kerr black hole. The intense frame-dragging in this region leads to the expectation that such emissions peak near the superradiant bound, at which the ratio of the angular momentum to the energy is the Schwarzschild radius. We solve analytically, to leading order in the deviation from the superradiant bound, the equations for null geodesics that connect emission points in the near-horizon region of an extreme Kerr to observation points in the region far from the black hole.\footnote{The $r$, $\phi$ and $t$ equations are solved generically while the $\theta$ equation is solved only in special cases.} Using geometric optics these are trajectories of photons which start near the horizon and end at the telescope. Previously such geodesics were known only numerically. 

These analytic results may be useful for the study of a variety of problems related to observations of electromagnetic radiation which originates from or passes through the near-horizon region of extreme  black holes. For example, we may use them to compute the profile of iron line emissions from the near-horizon of an extreme Kerr, which have been extensively analyzed in \cite{Laor:1991nc,Beckwith:2004nn,Dovciak:2004gq,Brenneman:2006hw,Dauser:2010ne}. We also hope to apply the equations derived herein to other potential observables such as extreme black holes silhouettes or orbiting hot spots.

The rest of the paper is organized as follows. In section \ref{Section: Geodesics} we set up the problem and give our conventions.  In section \ref{Section: r-theta} we solve for the radial null geodesic motion in the $r-\theta$ plane and in section \ref{Section: r-phi and r-t} we do the same for the motion in the $r-\phi$ and $r-t$ planes. Section \ref{Section: Conclusion} summarizes the results and discusses future directions.

\section{Geodesic equations}\label{Section: Geodesics}

The Kerr metric in Boyer-Lindquist coordinates $x^a\sim (\hat t,\hat r,\theta,\hat\phi)$ is given by:
\bea\label{Kerr}
ds^2&=&g_{ab}dx^adx^b\cr &=&-{\Delta \over\hat \rho^2}\left(d\hat t-a \sin^2\theta d\hat\phi\right)^2+{\sin^2 \theta \over \hat \rho^2}
\left((\hat r^2+a^2)d\hat \phi-a d\hat t\right)^2+{\hat\rho^2 \over\Delta}d\hat r^2+\hat \rho^2 d\theta^2\,,
\eea
\be
\Delta=\hat r^2-2M\hat{r}+a^2\,,\quad \hat \rho^2=\hat r^2 +a^2\cos^2 \theta \,. \notag
\ee
This describes neutral rotating black holes of mass $M$ and angular momentum $J=aM$.
The general form of the geodesic equation for $x^a(\tau)$ with affine parameter $\tau$ is 
\be 
\p_\tau^2 x^a+\Gamma^a_{bc}  \p_\tau x^b \p_\tau x^c=0\,, \quad 
g_{ab} \p_\tau x^a \p_\tau x^b =-\mu^2\,,
\ee   
where $\mu=0$ for null geodesics and $\mu=1$ for time-like geodesics. In the Kerr space-time there are three additional conserved quantities along every geodesic:
\bea 
\hat E&=&-g_{\hat t a}\p_\tau x^a\,,  \notag \\ 
\hat L&=&g_{\hat\phi a}\p_\tau x^a\,,\\ 
Q&=&Y_{ab}\p_\tau x^a \p_\tau x^b-(\hat L-a\hat E)^2\,. \notag
\eea
The Carter constant $Q$ here is constructed from the Killing tensor (square of Killing-Yano), 
\bea 
Y_{ab}d x^a d x^b&=&
a^2\cos^2\theta\left[\frac\Delta{\hat\rho^2}(d\hat t-a\sin^2\theta d\hat\phi)^2-\frac{\hat\rho^2}\Delta d\hat r^2\right]\notag\\
&&+\,\hat r^2\left[\frac{\sin^2\theta}{\hat\rho^2}\left(ad\hat t-(\hat r^2+a^2)d\hat\phi\right)^2+\hat\rho^2d\theta^2\right]\,. 
\eea
$\hat E$ and $\hat L$ are the energy and angular momentum while $Q$  measures the kinetic energy out of the equatorial plane. These conservation laws enable integration of the geodesic equation into the form \cite{Carter:1968rr}
%
\be\label{theta-r eqn hatted}
\int^{\hat{r}}{d\hat r'\over \sqrt{\hat R}}=\int^{\theta}{d\theta'\over \sqrt{\hat\Theta}}\,,
\ee 
\be\label{phi-r eqn hatted}
\hat{\phi}=\int^{\hat{r}}{a\hat{E}\hat{r'}^2+(\hat{L}-a\hat{E})(\Delta-a^2) \over \Delta\sqrt{\hat R}}d\hat r'+\int^{\theta}{\hat{L}\cot^2\theta' \over \sqrt{\hat\Theta}}d\theta'\,,
\ee
\be\label{t-r eqn hatted}
\hat{t}=\int^{\hat{r}}{\hat{E}\hat{r'}^2(\hat{r'}^2+a^2)+a(\hat{L}-a\hat{E})(\Delta-\hat{r'}^2-a^2)\over \Delta\sqrt{\hat R}}d\hat r'+\int^{\theta}{a^2\hat{E}\cos^2\theta' \over \sqrt{\hat\Theta}}d\theta'\,,
\ee
where
\bea
\hat{R}&=&\left(\hat E(\hat{r'}^2+a^2)-\hat L a\right)^2-\Delta\left(\mu^2 \hat{r'}^2+(\hat L-a\hat E)^2+Q\right)\,,\\
\hat\Theta&=&Q-\cos^2\theta'\left(a^2(\mu^2-\hat E^2)+ \hat{L}^2/\sin^2\theta'\right)\,.
\eea
The integrals are understood to be path integrals along the trajectory. At every point of the trajectory the conditions $\hat{R}\geq 0\,, \hat{\Theta}\geq 0$ must hold but the signs of $d\hat{r'}\,, d\theta'$ and $\sqrt{\hat R}\,, \sqrt{\hat \Theta}$ change at corresponding turning points in such a way that $d\hat r'/ \sqrt{\hat R}$ and $d\theta'/ \sqrt{\hat \Theta}$ are always positive. In general (\ref{theta-r eqn hatted}---\ref{t-r eqn hatted}) can only be evaluated numerically. 

In this paper we consider null geodesics,
\be 
\mu=0\,,
\ee
in the extreme Kerr space-time,
\be
a=M,
\ee
which are near the so-called superradiant bound:
\be \label{lambda defn}
\lambda=1-{\hat{L}\over 2M\hat{E}} \ll 1.
\ee
In this case we will find the radial integrals in (\ref{theta-r eqn hatted}---\ref{t-r eqn hatted}) are analytically soluble to leading order in $\lambda$. A particle or photon on a $\lambda <0$ geodesic cannot cross the horizon because, where it do so, the resulting black hole would violate the cosmic censorship bound. Hence, as already noted by Bardeen \cite{Bardeen:1973tla}, $\lambda=0$ is a special point in the parameter space of extreme Kerr geodesics. One expects many photons emitted from the near-horizon region to have small $\lambda$. It is therefore gratifying that this case is analytically soluble.

It is convenient to introduce the dimensionless Bardeen-Horowitz coordinates,
\be\label{NHEK coords}
t={\hat t \over 2M}\,,\quad \phi=\hat\phi-{\hat t\over 2M}\,,\quad r={\hat r-M\over M}\,. 
\ee
The near-horizon region is then $r\ll 1$, and the metric takes the form \cite{Bardeen:1999px}
\be\label{NHEK}
ds^2 = 2M^2 \Gamma(\theta)\left( -r^2 dt^2 + {dr^2 \over r^2} + d\theta^2 + \Lambda(\theta)^2(d\phi + r dt)^2\right)+\dots\,,
\ee
where $\Gamma(\theta) = {(1+\cos^2\theta)/ 2}\,, \Lambda(\theta) = {2\sin\theta/(1 + \cos^2\theta)}$
and the subleading corrections are suppressed by powers of $r$. The leading term displayed on the right hand side solves the Einstein equation on its own. It is referred to as the Near-Horizon Extreme Kerr or `NHEK'  metric  and has an enhanced symmetry group \cite{Bardeen:1999px,Guica:2008mu}.
We are interested in trajectories of photons emitted near the horizon and observed by a distant telescope. These are geodesics which start at $(t_n,r_n,\theta_n,\phi_n)$ with $r_n\ll 1$ in the NHEK region
and end at $(t_f,r_f,\theta_f,\phi_f)$ with $r_f\gg1$ in the far asymptotically flat region. 

In the null case  $\hat E$  may be scaled out of the geodesic equation, whose solutions are labeled by $\lambda$ and the convenient  dimensionless shifted Carter constant
\be\label{q defn}
\quad q^2=3-{Q\over M^2 \hat{E}^2}<4(1-\lambda+\lambda^2)\,,
\ee
where the last inequality expresses positivity of the kinetic energy in a local frame. 
We will show below that, given \eqref{lambda defn}, a geodesic that originates in NHEK and reaches out to the far asymptotically flat region must have  a positive $q^2$. We will take $q>0$ and hold it fixed while expanding in small $\lambda$. Motion in the equatorial plane has $q=\sqrt{3}$. 

In terms of the coordinates \eqref{NHEK coords} and parameters \eqref{lambda defn} and \eqref{q defn} the geodesic equations (\ref{theta-r eqn hatted}---\ref{t-r eqn hatted}) become:
\be\label{theta-r eqn}
\int_{r_n}^{r_f}{dr\over \sqrt{R}}=\int_{\theta_n}^{\theta_f}{d\theta\over \sqrt{\Theta}}\,,
\ee 
\be\label{phi-r eqn}
\phi_f-\phi_n=-\half\int_{r_n}^{r_f}{\Phi \over r\sqrt{R}}dr 
+\half\int_{\theta_n}^{\theta_f}{(3+\cos^2\theta-4\lambda)\cot^2\theta\over \sqrt{\Theta}}d\theta\,,
\ee
\be\label{t-r eqn}
t_f-t_n=\half\int_{r_n}^{r_f}{T\over r^2\sqrt{R}}dr +\half\int_{\theta_n}^{\theta_f}{\cos^2\theta \over \sqrt{\Theta}}d\theta\,,
\ee
where 
\bea
R&=&r^4+4r^3+(q^2+8\lambda-4\lambda^2)r^2+8\lambda r+4\lambda^2\,,\\
\Theta&=&3-q^2+\cos^2\theta-4(1-\lambda)^2\cot^2\theta\,,\\
\Phi&=&r^3+4r^2+(3+4\lambda)r+4\lambda\,,\\
T&=&r^4+4r^3+7r^2+4(1+\lambda)r+4\lambda\,.
\eea

The integrals in (\ref{theta-r eqn}---\ref{t-r eqn}) are of elliptic type and are generally treated numerically. In this paper we will perform the radial integrals analytically to leading order in small $\lambda$ using the method of matched asymptotic expansions (MAE) for all geodesics  extending from the near to the far region.  We will proceed in the small $\lambda$ regime by dividing the spacetime into two regions: \be\label{nr} {\bf Near~~Region} ~~~~~r\ll 1\,,\ee  \be \label{fr}{\bf Far~~Region} ~~~~~r\gg\sqrt{\lambda}\,.\ee  The two regions overlap in the  \be \label{or} {\bf Overlap~~Region} ~~~~~\sqrt{\lambda} \ll r\ll 1\,.\ee We now proceed to solve the equations in the near and far regions and match the solutions in the overlap region. 

\section{The $r$--$\theta$ motion}\label{Section: r-theta}
In this section we solve the radial integral 
\be\label{I(r) theta}
I=\int_{r_n}^{r_f} \frac{dr}{\sqrt{R(r)}}\,,
\ee to leading order in $\lambda$ via MAE. 
Given \eqref{lambda defn} we have that
\be
R\approx r^4+4r^3+q^2r^2+8\lambda r+4\lambda^2\,.
\ee
In the near region (\ref{nr}) we have
\be
R_{n}(r)=q^2r^2+8\lambda r+4\lambda^2\,,
\ee
while in the far region (\ref{fr}) we have
\be
R_{f}(r)=r^2(r^2+4r+q^2)\,.
\ee

We first check the conditions under which small $\lambda$ geodesics in the near and far regions connect. Turning points in the radial motion occur at the zeroes of $R(r)$. In the far region these are located at 
\be r_{f\pm}=-2\pm\sqrt{4-q^2}\,.\ee
If $q^2$ is positive, this has no turning points with positive $r$. On the other hand, if $q^2$ is negative, there is a turning point  at  positive $r$, and the geodesic bounces off the black hole before it penetrates the near region. We therefore take 
\be \label{q positive}
q^2>0\,,
\ee
ignoring the measure zero case $q=0$. In the near region the zeroes are at 
\be r_{n\pm}={2\lambda  \over  q^2}\left(-2 \pm \sqrt{4-q^2}\right)\,.\ee
Since $0<q^2<4$ to leading order in $\lambda$, the square root is always a positive number less than 2. Hence there is a positive root if and only if
$\lambda$ is negative. In that case, a geodesic emanating from the horizon will turn around before it reaches the far region at $r\sim \sqrt{\lambda}$. This is due to the fact that photons on such trajectories exceed the superradiant bound. 
On the other hand, there are two turning points, and if the geodesic originates in the near region but outside both turning points, $r>r_{n\pm}$, it can reach the far region. Hence $\lambda$ can have either sign but geodesics with negative $\lambda$ cannot get all the way to the horizon.

To leading order in $\lambda$ the near and far radial integrals can now be performed analytically:
\bea
I_{n}(r)&=&\int^r \frac{dr'}{\sqrt{R_{n}(r')}} = {1\over q}\ln\left(q\sqrt{R_{n}(r)}+q^2 r+4\lambda \right) +C_{n} \,, \label{near soln theta} \\
I_{f}(r)&=&\int^r \frac{dr'}{\sqrt{R_{f}(r')}} = -{1\over q} \ln{1\over r^2} \left(q\sqrt{R_{f}(r)} +q^2r+2r^2\right) +C_{f}\,, \label{far soln theta}
\eea
where $C_{n}, C_{f}$ are integration constants.
In the overlap region (\ref{or}) we have:
\bea 
I_{n}(r)&=&{1\over q}\left(\ln r+\ln 2q^2+q\,C_{n}+{4\lambda\over q^2 r}+\ldots\right)\,,\\
I_{f}(r)&=&{1\over q} \left(\ln r-\ln 2q^2+q\,C_{f}-{2r\over q^2}+\ldots\right)\,.
\eea
Matching $I_{n}=I_{f}$ in the overlap region we find:
\be\label{matching constants theta}
C_{f}=C_{n}+{2\over q}\ln 2q^2\,.
\ee
The integral \eqref{I(r) theta} is given by $I=I_f(r_f)-I_n(r_n)$ and using \eqref{near soln theta}, \eqref{far soln theta}, and \eqref{matching constants theta} we find:
\be\label{r-theta soln}
I=-{1\over q}\ln{\left(q D_{n}+q^2 r_n+4\lambda\right)\left(q D_{f}+2r_f+q^2\right)\over 4q^4\, r_f}\,,
\ee
where,
\be\label{Dn Df def}
D_{n}=\sqrt{R_{n}(r_n)}=\sqrt{q^2 r_n^2+8\lambda r_n+4\lambda^2}\,,\quad D_{f}={1\over r_f}\sqrt{R_{f}(r_f)}=\sqrt{r_f^2+4r_f+q^2}\,.
\ee

\section{The $r$--$\phi$ and $r$--$t$ motion}\label{Section: r-phi and r-t}

Given \eqref{lambda defn} we have that
\bea
\Phi&\approx&r^3+4r^2+3r+4\lambda\,,\\
T&\approx&r^4+4r^3+7r^2+4r+4\lambda\,,
\eea
and we can perform the radial integrals for the $r$--$\phi$ and $r$--$t$ motion in \eqref{phi-r eqn} and \eqref{t-r eqn},
\be\label{I(r) phi and t}
I^\phi =\int_{r_n}^{r_f} \frac{\Phi(r)}{r\sqrt{R(r)}}dr\,,\quad 
I^t =\int_{r_n}^{r_f} \frac{T(r)}{r^2\sqrt{R(r)}}dr\,,
\ee 
via MAE as follows.

In the near region \eqref{nr} we have
\be
\Phi_{n}(r)=3r+4\lambda\,,\quad T_{n}(r)=4r+4\lambda\,,
\ee
while in the far region \eqref{fr} we have
\be
\Phi_{f}(r)=r(r^2+4r+3)\,,\quad T_{f}(r)=r(r^3+4r^2+7r+4)\,.
\ee
All integrals are now doable:
\bea
I^\phi_{n}(r)=\int^r \frac{\Phi_{n}(r')}{r'\sqrt{R_{n}(r')}}dr'
&=&{3\over q}\ln\left(q\sqrt{R_{n}(r)}+q^2 r+4\lambda\right)- \label{near soln phi} \\ 
&&-2\ln{1\over r}\left(\sqrt{R_{n}(r)}+ 2r+2\lambda\right) +C^{\phi}_{n}  \,,  \notag \\
I^\phi_{f}(r)=\int^r \frac{\Phi_{f}(r')}{r'\sqrt{R_{f}(r')}}dr' 
&=&-{3\over q}\ln{1\over r^2} \left(q\sqrt{R_{f}(r)}+q^2r+2r^2\right)+ \label{far soln phi} \\ &&+2\ln{1\over r}\left(\sqrt{R_{f}(r)}+r^2+2r\right)+ {1\over r}\sqrt{R_{f}(r)} +C^{\phi}_{f}\,,  \notag \\
I^t_{n}(r)=\int^r \frac{T_{n}(r')}{{r'}^2\sqrt{R_{n}(r')}}dr' 
&=& -{1\over \lambda r} \sqrt{R_{n}(r)} +C^{t}_{n}\,, \label{near soln t} \\
I^t_{f}(r)=\int^r \frac{T_{f}(r')}{{r'}^2\sqrt{R_{f}(r')}}dr' 
&=& -{7q^2-8\over q^3} \ln{1\over r^2}\left(q\sqrt{R_{f}(r)}+q^2r+2r^2\right)+ \label{far soln t} \\ 
&& +2\ln{1\over r}\left(\sqrt{R_{f}(r)}+r^2+2r\right) +{q^2r-4\over q^2r^2}\sqrt{R_{f}(r)} +C^{t}_{f}\,,  \notag
\eea
where $C^{\phi,t}_{n}, C^{\phi,t}_{f}$ are integration constants.
In the overlap region \eqref{or} we have:
\bea
I^\phi_{n}(r)&=&{1\over q}\left(3\ln r+3\ln 2q^2+q\left(C^\phi_{n}-2\ln(q+2)\right)-{4\lambda(q^2-3)\over q^2 r}+\ldots\right)\,,\\
I^\phi_{f}(r)&=&{1\over q}\bigg(3\ln r-3\ln 2q^2 +q\left(C^\phi_{f} +2\ln(q+2)\right)
+q^2+{2(2q^2-3)r\over q^2}+\ldots\bigg)\,,\\ 
I^t_{n}(r)&=&-{1\over q}\left({q^2\over \lambda}-q\,C^t_{n}+{4\over r}+{2\lambda(q^2-4)\over q^2 r^2}+\ldots\right)\,,\\
I^t_{f}(r)&=&-{1\over q}\left({4\over r}-{7q^2-8\over q^2}\ln r -{q^4-(7q^2-8)\ln 2q^2-8\over q^2}- \right.\\
&&\qquad\quad \left.-q\left(C^t_{f}+2\ln(q+2)\right)-{4(q^4-4q^2+6)r\over q^4}+\ldots\right)\,, \notag
\eea
so matching $I^{\phi,t}_{n} =I^{\phi,t}_{f}$ in the overlap region we find:
\bea
C^\phi_{f}&=&C^\phi_{n}+{6\over q}\ln 2q^2 -4\ln(q+2)-q \,, \label{matching constants phi} \\
C^t_{f}&=&C^t_{n}-{q\over\lambda}- \frac{q^4-(7q^2-8)\ln 2q^2-8}{q^3}-2\ln(q+2)\approx C^t_{n}-{q\over\lambda}\,, \label{matching constants t}
\eea
where in the last line we have used \eqref{lambda defn}.
The integrals \eqref{I(r) phi and t} are given by  $I^{\phi,t}=I^{\phi,t}_f(r_f)-I^{\phi,t}_n(r_n)$ and using (\ref{near soln phi}---\ref{far soln t}) and (\ref{matching constants phi}---\ref{matching constants t}) we find:
\bea
I^{\phi}&=&-{3\over q}\ln{1\over r_f}\left(qD_{n}+ q^2 r_n+ 4\lambda\right)\left(qD_{f}+ 2r_f+q^2\right) \label{r-phi soln} \\
&&+2\ln{1\over r_n} \left(D_{n}+2r_n+2\lambda\right) \left(D_{f}+r_f+2\right)+D_{f}+{6 \over q}\ln 2q^2 -4\ln(q+2)-q\,,\notag \\ 
I^{t}&=&2\ln\left(D_{f}+r_f+2 \right)-{7q^2-8\over q^{3}}\ln{1\over r_f} \left(qD_{f}+2r_f+q^2\right) \label{r-t soln} \\
&&+{q^2r_f-4 \over  q^2r_f}D_{f} +{1 \over  \lambda r_n}D_{n}-{q\over  \lambda}\,.\notag
\eea
In the special case of motion confined to the equatorial plane the geodesic equations reduce to \eqref{phi-r eqn} and \eqref{t-r eqn} with the $\theta$ integrals dropped. In this case the shifts in the azimuthal angle $\phi$ and time $t$ are given by $\phi_f-\phi_n=-(1/2)\,I^{\phi}\,\vert_{q=\sqrt{3}}$ and $ t_f-t_n=(1/2)\,I^{t}\,\vert_{q=\sqrt{3}}$ respectively, or explicitly:
\bea
\phi_f-\phi_n&=&{\sqrt{3}\over 2}\ln{1\over r_f}\left(\sqrt{3}\,\mathcal{D}_{n}+ 3 r_n+ 4\lambda\right)\left(\sqrt{3}\,\mathcal{D}_{f}+ 2r_f+3\right) \label{theta=cnst phi soln} \\
&&-\ln{1\over r_n} \left(\mathcal{D}_{n}+2r_n+2\lambda\right) \left(\mathcal{D}_{f}+r_f+2\right)-\half\mathcal{D}_{f}-\sqrt{3}\ln 6 +2\ln(\sqrt{3}+2)+{\sqrt{3}\over 2}\,,\notag \\ 
t_f-t_n&=&\ln\left(\mathcal{D}_{f}+r_f+2 \right)-{13\over 6\sqrt{3}}\ln{1\over r_f} \left(\sqrt{3}\,\mathcal{D}_{f}+2r_f+3\right) \label{theta=cnst t soln} \\
&&+{3r_f-4 \over  6r_f}\mathcal{D}_{f} +{1 \over 2 \lambda r_n}\mathcal{D}_{n}-{\sqrt{3}\over 2 \lambda}\,,\notag
\eea
with $\mathcal{D}_n=D_n\vert_{q=\sqrt{3}}=\sqrt{3 r_n^2+8\lambda r_n+4\lambda^2}\,, \mathcal{D}_f=D_f\vert_{q=\sqrt{3}}=\sqrt{r_f^2+4r_f+3}$. These are algebraic equations which relate an endpoint of an equatorial null geodesic near the horizon $(t_n,r_n,\phi_n)$ to its endpoint in the region far from the black hole $(t_f,r_f,\phi_f)$ and the associated constant $\lambda$ along the geodesic.


\section{Conclusion}\label{Section: Conclusion}
In this paper we studied null geodesics in the extreme Kerr space-time. 
We used the method of matched asymptotic expansions to integrate, to leading order in the deviation from the superradiant bound, all radial integrals in Carter's integral geodesic equations. 
The key equations derived in this paper are \eqref{r-theta soln}, \eqref{r-phi soln}, and \eqref{r-t soln}. 

In the special case of motion confined to the equatorial plane we derived the algebraic equations (\ref{theta=cnst phi soln},\ref{theta=cnst t soln}) which relate an endpoint of a null geodesic near the horizon to its endpoint in the region far from the black hole  and the associated angular momentum constant along the geodesic. 

For geodesics which move in the $\theta$ direction we note that positivity of the shifted Carter constant $q^2$ (derived in \eqref{q positive} for all geodesics which connect between the near and far regions) restricts the values of $\theta$ that these geodesics may possibly explore. This is because $\Theta\geq 0$ then implies $3+\cos^2\theta-4\cot^2\theta\geq 0$ which means that $\theta$ must lie between $\theta_0=\arccos\sqrt{2\sqrt{3}-3}\approx 47^\circ$ and $\pi-\theta_0\approx 133^\circ$. This fact has been derived previously from an analysis of the NHEK geodesics alone in \cite{AlZahrani:2010qb}.

Solving the null geodesic equations studied in this paper for a fixed observer far from the black hole and using the geometrical optics methods of \cite{Cunningham-Bardeen(1973),Cunningham(1975)} one may obtain various observables related to propagation of electromagnetic radiation from the near-horizon region of extreme black holes. These include broadened Fe K$\alpha$ emission lines and images of hot orbiting spots. We hope that the results obtained in this paper together with an analytical treatment of the $\theta$ integrals will lead to a derivation of analytical formulas for such observables.

Finally, it is worth mentioning that while in our expansion in small $\lambda$ we have been keeping $q^2$ fixed and finite, all of the equations and statements derived in this paper are also true for the case of small $q^2$ provided that it remains $q^2\gg\sqrt{\lambda}$. In this case the only thing that needs to be modified are the definitions of the near region \eqref{nr} to $r\ll \textrm{min}\left(1,q^2\right)$ and the overlap region \eqref{or} to $\sqrt{\lambda} \ll r\ll \textrm{min}\left(1,q^2\right)$.

\section*{Acknowledgements}

We are grateful to Sam Gralla and Alex Lupsasca for useful conversations. This work was supported in part by NSF grant 1205550, Templeton foundation award 52476, and the Sir Keith Murdoch Fellowship.

\end{document}